\documentclass[aps,prb,twocolumn,superscriptaddress,showpacs]{revtex4}
\usepackage{graphicx}
\usepackage{dcolumn}
\usepackage{bm}

\usepackage{color}

\begin{document}


\title{Large Seebeck effect in the electron-doped FeAs$_2$ driven by 
quasi one dimensional pudding-mold type band}


\author{Hidetomo Usui}
\affiliation{Department of Engineering Science,
The University of Electro-Communication, Chofu, Tokyo 182-8585, Japan}
\affiliation{JST, ALCA, Gobancho, Chiyoda, Tokyo 102-0076, Japan}

\author{Kazuhiko Kuroki}
\affiliation{Department of Engineering Science,
The University of Electro-Communication, Chofu, Tokyo 182-8585, Japan}
\affiliation{JST, ALCA, Gobancho, Chiyoda, Tokyo 102-0076, Japan}

\author{Seiya Nakano}
\affiliation{Department of Physics, Okayama University, Okayama 
700-8530, Japan}
\affiliation{JST, ALCA, Gobancho, Chiyoda, Tokyo 102-0076, Japan}
\author{Kazutaka Kudo}
\affiliation{Department of Physics, Okayama University, Okayama 
700-8530, Japan}
\affiliation{JST, ALCA, Gobancho, Chiyoda, Tokyo 102-0076, Japan}
\author{Minoru Nohara}
\affiliation{Department of Physics, Okayama University, Okayama 
700-8530, Japan}
\affiliation{JST, ALCA, Gobancho, Chiyoda, Tokyo 102-0076, Japan}


\date{\today}

\begin{abstract}
We investigate the thermoelectric propeties of the electron-doped 
FeAs$_2$ both experimentally and theoretically. Electrons are 
doped by partially substituting Se for As, which leads to a metallic 
behavior in the resistivity. A Seebeck coefficient of about $-$200 $\mu$V/K 
is reached at 300 K for 1\% doping, and  about $-$120 $\mu$V/K 
even at 5\% doping. 
The origin of this large Seebeck coefficient despite the 
metallic conductivity is analyzed from a band 
structure point of view. The first-principles band calculation 
reveals the presence of a pudding-mold type band just above the band gap, 
somewhat similar to Na$_x$CoO$_2$, but with a quasi-one-dimensional nature.
We calculate the Seebeck coefficient using a tightbinding model 
that correctly reproduces this band structure, and this 
gives results close to the experimental 
observations. The origin of this peculiar band shape is also discussed.
\end{abstract}


\maketitle

\section{Introduction}
Searching materials with good thermoelectric properties has 
been an issue of great interest both from the viewpoint of 
fundamental physics and application purposes.
The efficiency of a thermoelectric material is often evaluated by 
the dimensionless figure of merit $ZT=S^2 T/(\rho\kappa)$, where 
$S$ is the Seebeck coefficient, $\rho$ the electrical resistivity, 
$\kappa$ the thermal conductivity, and $T$ being the temperature\cite{Mahan}.
$S^2/\rho$ in the $ZT$ formula is called the power factor, 
and in order to have this quantity large,  
both large Seebeck coefficient and small resistivity are required.
However, it is usually difficult to have large Seebeck effect in 
metallic materials, so 
good thermoelectric materials are often found in narrow gap 
semiconductors. 

The discovery of the large Seebeck effect in Na$_x$CoO$_2$\cite{Terasaki} 
has provided a new avenue for searching good thermoelectric 
properties in materials with metallic resistivity with significantly 
amount of carriers.
In fact, such a coexistence of metallic resistivity and 
large Seebeck coefficient have also been 
observed in e.g., CuRhO$_2$\cite{Kuriyama,Shibasaki2}, 
LiRh$_2$O$_4$\cite{Okamoto}, LaRhO$_3$\cite{Shibasaki1}, 
SrTiO$_3$\cite{Okuda,Ohta}, 
and quite recently, in PtSb$_2$\cite{Nishikubo}.
Theoretical studies on the origin of the large Seebeck 
effect in these materials have also been extensively performed
\cite{Singh,Wilson,Maekawa,Kuroki,Usui,Arita,Usui_SrTiO3,Held,Mori}.
In particular, Kuroki and Arita pointed out in Ref.~\onlinecite{Kuroki} that a 
peculiar band shape referred to as the ``pudding-mold'' type, which consists of a dispersive portion and a flat portion, can 
in general be favorable for the coexistence of large Seebeck coefficient 
and small resistivity. The quasi-two-dimensional pudding mold band 
has been proposed as an origin of the large power factor observed 
in Na$_x$CoO$_2$ and other materials. 
On the other hand, Mori {\it et al.} have recently 
studied the origin of the large Seebeck effect in 
PtSb$_2$, and found a three dimensional ``corrugated'' 
flat band that gives rise to many Fermi surface pockets through the 
entire Brillouin zone when holes are doped.\cite{Mori} This 
also gives rise to a coexisting large Seebeck coefficient and metallic 
resistivity in general.

In the present study, we focus on FeAs$_2$
in an attempt to look for yet another material that 
exhibits large Seebeck effect despite the metallic resistivity. 
FeAs$_2$ crystallizes in an orthorhombic Marcasite structure with {\it Pnnm} space group (No. 58). 
The structure is characterized by the edge-sharing FeAs$_6$ octahedra, which are connected along the 
crystallographic $c$ axis ($z$ axis) to form linear chains.
The non-doped semiconductor FeAs$_2$
has been investigated previously in Ref.~\onlinecite{PSun} in comparison with 
FeSb$_2$, which exhibits a colossal thermoelectric effect\cite{Bentien}. 
A theoretical analysis has shown  that the Seebeck coefficient of 
the non-doped FeAs$_2$ can be understood quantitatively using the 
band structure obtained from the first-principles calculation\cite{Kotliar}.
In the present study, we investigate the thermoelectric properties 
of the {\it electron doped} FeAs$_2$ both experimentally and theoretically.
By doping electrons by partially substituting Se for As, 
the resistivity exhibits a metallic behavior, but nonetheless we observe 
a fairly large Seebeck coefficient at 300 K of about 
$-$200$ \mu$V/K at 1\% doping and $-$120 $\mu$V/K even at 5\%
doping. 
We also analyze the origin of this experimental observation 
theoretically. It is found that the quasi-one-dimensional (q1D) 
version of the pudding-mold type band is present right above the 
band gap, and this peculiar band shape gives rise to the 
coexistence of the metallic resistivity and the large 
Seebeck coefficient.  We also discuss the origin of this peculiar 
band shape using a simplified model. 
With the present study, we now have examples of materials 
that possess one- (FeAs$_2$), 
two- (Na$_x$CoO$_2$, etc.), or three-dimensional (PtSb$_2$) 
flat bands that give rise to the large 
Seebeck coefficient with the metallic resistivity.
This reinforces the general efficiency of peculiar band shapes with 
partially flat dispersions.

\section{Experiment}

\subsection{Synthesis}
Polycrystalline samples of Fe(As$_{1-x}$Se$_x$)$_2$ with nominal Se contents of $x$ = 0.0, 0.01, 0.025, and 0.05 were synthesized by a solid-state reaction in two steps. 
First, stoichiometric amount of starting materials Fe (99.9\%), As (99.9999\%), and Se (99.9\%) were mixed and ground. 
They were heated in an evacuated quartz tube at 500$^{\circ}$C for 40 h and then at 700$^{\circ}$C for 40 h. 
Second, the product was powdered, pressed into pellets, and sintered at 800$^{\circ}$C for 12 h. 
The obtained samples were characterized by powder X-ray diffraction (XRD) and confirmed to be a single phase of Fe(As$_{1-x}$Se$_x$)$_2$. 
Electrical resistivity $\rho$ and Seebeck coefficient $S$ were measured using a Physical Property Measurement System (PPMS, Quantum Design) 
in the temperature range from 2 to 300 K.

\subsection{Thermoelectric properties}
\begin{figure}
\includegraphics[width=7.5cm]{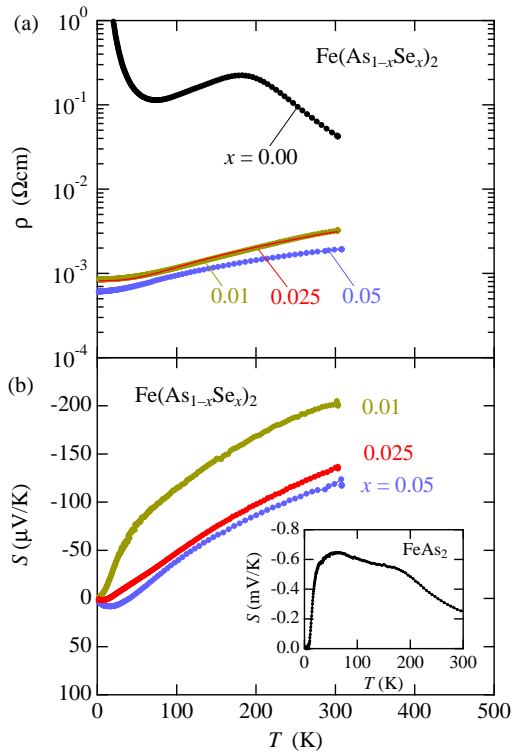}
\caption{
Temperature dependence of (a) electrical resistivity $\rho(T)$ and Seebeck coefficient $S(T)$ of polycrystalline Fe(As$_{1-x}$Se$_x$)$_2$ with $x$ = 0.00, 0.01, 0.025, and 0.05. The inset shows Seebeck coefficient of non-doped FeAs$_2$. 
\label{fig_exp1}
}
\end{figure}

As shown in Fig.~\ref{fig_exp1}(a), non-doped FeAs$_2$ exhibits a semiconducting behavior, consistent with the previous reports.~\cite{PSun}
The Seebeck coefficient $S$ of the non-doped FeAs$_2$ exhibits a large maximum value of $-$0.65 mV/K at approximately 60 K, as shown in 
the inset of Fig.~\ref{fig_exp1}(b), which is similar to the literature data.~\cite{PSun}
Then, both the magnitude and temperature dependence of resistivity $\rho$ changed abruptly by substituting Se for As from semiconducting 
to metallic.  The $\rho$ values on the order of 1 m$\Omega$cm and the positive temperature slope of $\rho$ 
suggest that a metallic sate is realized for Fe(As$_{1-x}$Se$_x$)$_2$ ($x$ = 0.01, 0.025, and 0.05). 
The temperature dependence of $S$ is also changed abruptly by Se doping. 
But even in the metallic state,  the Seebeck coefficient reaches a fairly large value of about $-$200 $\mu$V/K at 300 K for the 
$x = 0.01$ sample, as shown in Fig.~\ref{fig_exp1}(b). 
In this way, metallic $\rho$ is compatible with a large $S$ for Fe(As$_{1-x}$Se$_x$)$_2$. 
This combination of the resistivity and the Seebeck coefficient gives a maximum power factor of 14 $\mu$W/cm K$^2$ 
at approximately 200 K for the $x$ = 0.01 sample.

\section{Theoretical Analysis}
\subsection{Band structure}

We now turn to the theoretical analysis.
We perform first-principles band structure calculation using 
the Wien2k package\cite{Wien2k}. 
The structural parameters adopted are given in Ref.~\onlinecite{FeAs2_structure},  
in which the lattice constants are $a=5.2684, b=5.9631$, and $c=2.9007$ ${\rm\AA}$, 
space group is {\it Pnnm}, and the atomic positions are 
$(0,0,0)$ for Fe and $(0.1760,0.3625,0)$ for As.
In the present study, we construct tightbinding models 
in order to calculate the Seebeck coefficient.
To obtain a model that correctly reproduces the first-principles band structure,
we construct maximally localized wannier functions (MLWFs)
from the first-principles calculation result.\cite{wannier90}
We focus on the electron doped case up to 
about the room temperature, 
so that we only have to consider the band structure above the band gap.
The tight-binding Hamiltonian is written as
\begin{eqnarray}
H = \sum_{ij\sigma}\sum_{\mu\nu}t^{\mu\nu}_{ij}c^{\dagger}_{i\mu\sigma}c_{j\nu\sigma},
\end{eqnarray}
where $i,j$ denote the sites, and $\mu,\nu$ the orbitals,  
and $t^{\mu\nu}_{ij}$ is the 
hopping parameter obtained from MLWFs , 
and $c^{\dagger},c$ are the creation-annihilation operators.
Here we first consider six orbitals (three orbitals per Fe) 
to reproduce the bands up to about 5eV above the Fermi energy.
These orbitals have mainly $3d$ character, although they are 
hybridized with As $4p$.

\begin{figure}
\includegraphics[width=8.5cm]{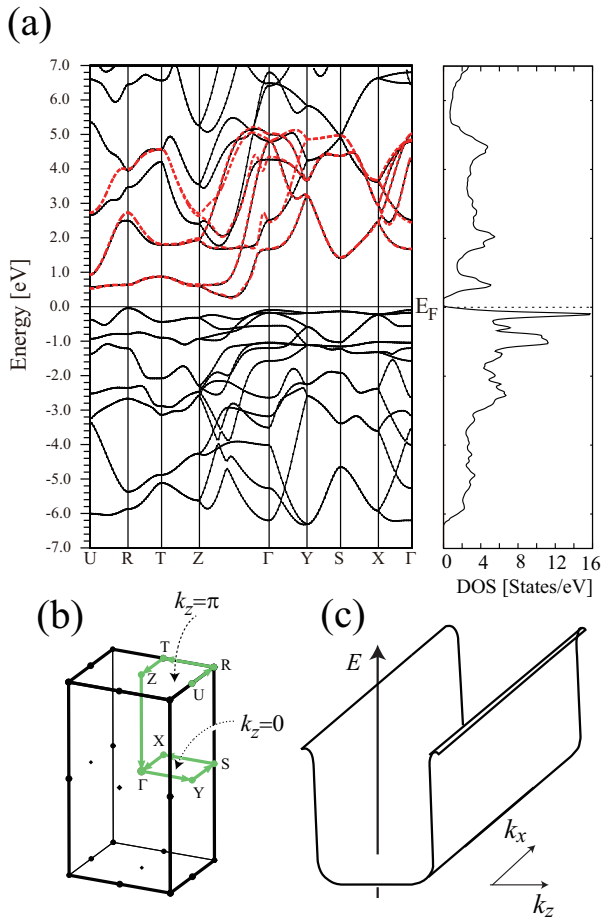}
\caption{(a) The band structure and the density of states of FeAs$_2$. 
The solid black lines are the original first-principles band structure, 
while the red dashed lines are the bands of the tight-binding model 
constructed from MLWFs. 
(b) The Brillouin zone is shown.  (c) A schematic image of the q1D 
pudding-mold type band of FeAs$_2$.
\label{fig_theory1}
}
\end{figure}

\begin{figure}
\includegraphics[width=8.5cm]{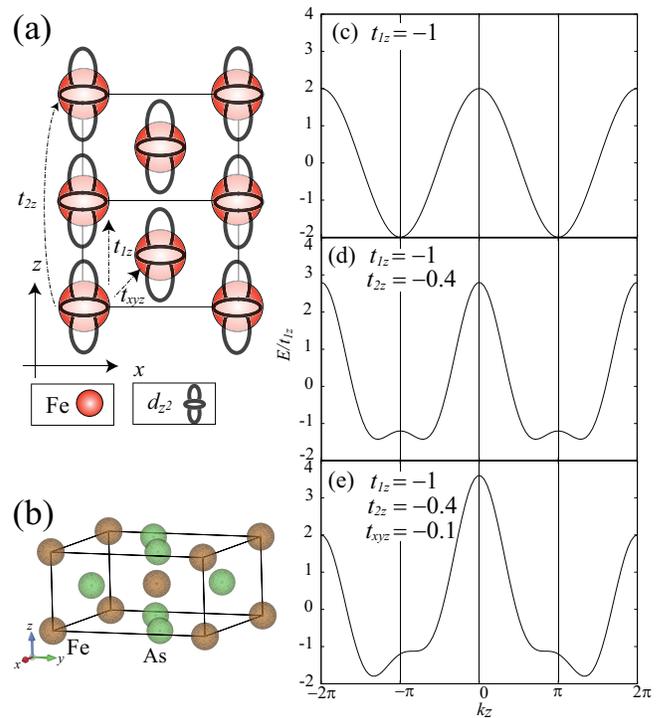}
\caption{(a) A simplified tightbinding model of FeAs$_2$ and 
(b) crystal structure of FeAs$_2$.
The band structure in the unfolded Brillouin zone 
is shown in the right panel with (c)$t_{1z}=-1$, 
(d)$t_{1z}=-1$ and $t_{2z}=-0.4$, and 
(e)$t_{1z}=-1$, $t_{2z}=-0.4$ and $t_{xyz}=-0.1$.
\label{fig_theory4}
}
\end{figure}

The first-principles band structure is shown in Fig.\ref{fig_theory1}(a)  
with the tight binding bands superposed.
The band structure just above the band gap 
has a nearly flat dispersion along U-R-T-Z (see (b) for the 
symmetrical points in the Brillouin zone), namely, 
within the $k_z=\pi$ plane. 
In fact, the Wannier orbitals that are the origin of these bands 
are found to be elongated in the $z$ direction
(see Fig.\ref{fig_theory4}).
We will call these orbitals the ``$d_{z^2}$ orbitals'' for simplicity
(the maximally localized Wannier orbitals are obtained 
by projecting onto the $d_{z^2}$ orbital), 
although they 
are actually more complicated due to the hybridization with the 
As 4p orbitals.
The quasi-one-dimensionality of the bands within the $k_z=\pi$ plane 
is partially because of the anisotropy of these $d_{z^2}$ orbitals, 
but to be more precise, there is another factor that restricts this 
quasi-one-dimensionality 
to near the $k_z=\pi$ plane, which will be discussed later.
Furthermore, these bands are somewhat flat also 
along Z-$\Gamma$ (i.e., the $k_z$ direction) 
up to some intermediate point, then bend sharply into dispersive portions.
In Fig.\ref{fig_theory1}(c), we show a schematic figure of this band structure.
It has a flat portion at the bottom with a q1D nature,
so it is a q1D version of the pudding-mold type band introduced in 
ref.\onlinecite{Kuroki} as the band structure of Na$_x$CoO$_2$.
This type of band structure is favorable for producing good 
thermoelectric properties when carriers are doped. The reason is because 
(i) even when a large amount of carriers are doped, the Fermi level 
sticks close to the band bottom due to the large density of states,
(ii) for such a position of the Fermi level, there is a large 
difference between electron and hole group velocity, resulting 
in a large Seebeck coefficient, and (iii) small resistivity 
can be achieved owing to the large amount of doped carriers and the 
dispersive portion of the band near the Fermi level.
The origin of such a peculiar band structure in FeAs$_2$ will be 
discussed later.

\begin{figure}
\includegraphics[width=8.5cm]{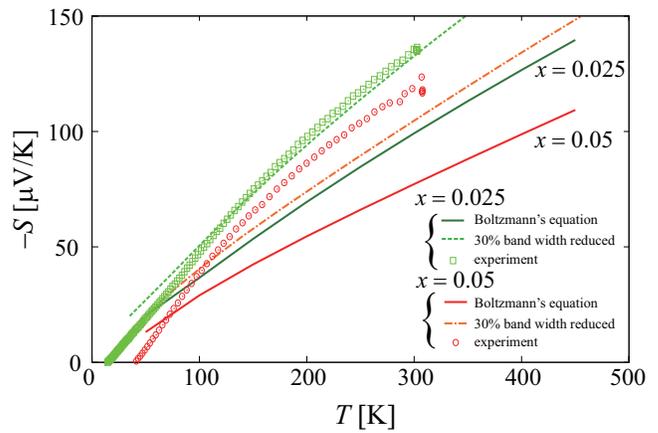}
\caption{The calculated Seebeck coefficient for $x=0.025$ and $x=0.05$ 
(the solid lines)  compared with the experimental result 
(green squares and red circles). A calculation result with the band 
width reduced by 30$\%$ (green dotted line and red dash-dotted line).
\label{fig_theory2}
}
\end{figure}

\subsection{Seebeck coefficient}
Let us now move on to the Seebeck coefficient calculation.
We first briefly introduce the method to 
calculate the Seebeck coefficient and the power factor.\cite{Wilson,Mahan}
The Seebeck coefficient $S$ and the resistivity $\rho$ are 
calculated with Boltzmann's equation 
using the tight-binding Hamiltonian.
They are given as, 
\begin{eqnarray}
{\bf S} = \frac{1}{eT}\frac{{\bf K}_1}{{\bf K}_0}, \\
\bm{\rho} = e^{-2}{\bf K}^{-1}_0,
\end{eqnarray}
where $e$ is the elementary electrical charge($e<0$), 
$T$ is the temperature. ${\bf K}_0$ and ${\bf K}_1$ are given as
\begin{eqnarray}
{\bf K}_n = \sum_{{\bf k}\sigma \nu} \tau({\bf k}){\bf v}^{\nu}_{k}{\bf v}^{\nu}_{k}\left(-\frac{\partial f(\varepsilon)}{\partial \varepsilon}({\bf k})\right)(\varepsilon^{\nu}_{\bf k} - \mu)^n,
\end{eqnarray}
where ${\bf k}$ is wave vector, $\tau$ is quasi-particle lifetime, 
$\varepsilon^{\nu}_{\bf k}$ is the $\nu$-th energy eigenvalue at ${\bf k}$, 
${\bf v}^{\nu}_{\bf k}=(1/\hbar)(d\varepsilon^{\nu}_{\bf k}/d{\bf k})$ is the group velocity tensor,
$f$ is the Fermi distribution function, and $\mu$ is the chemical potential at $T$.
We simply denote $(K_n)_{zz} = K_n$, $S_{zz} = S = (1/eT)(K_0/K_1)$,  
and $\rho_{zz} = \rho = e^{-2}K_{0}^{-1}$.
We will approximate the quasiparticle life time as an (undetermined) 
constant in the present study, so that it cancels out 
in the Seebeck coefficient calculation, while not in the resistivity. 
Thus the absolute value of the power factor $S^2/\rho$ is not 
determined and will be normalized by a certain value.

The calculated Seebeck coefficient as a function of 
the temperature is shown in Fig.\ref{fig_theory2}.
The results at $T=300$K give $-100\mu$V/K at $x=0.025$, 
and $-77\mu$V/K at $x=0.05$.
The theoretical result is about 20\% smaller 
than the experimental values at $x=0.025$ and $T=300$K.
This discrepancy from the experiment may be due to 
the constant lifetime approximation, or the band width renormalization 
due to the electron correlation effects that are not considered in the 
first-principles band calculation.
As for the latter possibility, 
it has been known from the comparison between the 
band calculation and the angle-resolved photoemission studies 
that the bandwidth of the $3d$ electron materials can be 
reduced by as much as 60\% , and taking this effect into account 
in the Seebeck coefficient calculation 
reproduces quantitatively 
the experimental results of Na$_x$CoO$_2$\cite{Kuroki}.
The Seebeck coefficient is underestimated also in other cases of $3d$ electron 
systems such as SrTiO$_3$\cite{Usui_SrTiO3}, due to the band 
width renormalization  effect.
If we assume that the band width is reduced by 30\% from the first-principles 
result,  the calculated Seebeck coefficient is in good agreement with the 
experimental result as shown in Fig.\ref{fig_theory2}(dotted line). 
The band width renormalization effect is not so strong as in Na$_x$CoO$_2$,
indicating that the electron correlation effects are not so large in this 
material.

\begin{figure}
\includegraphics[width=6.5cm]{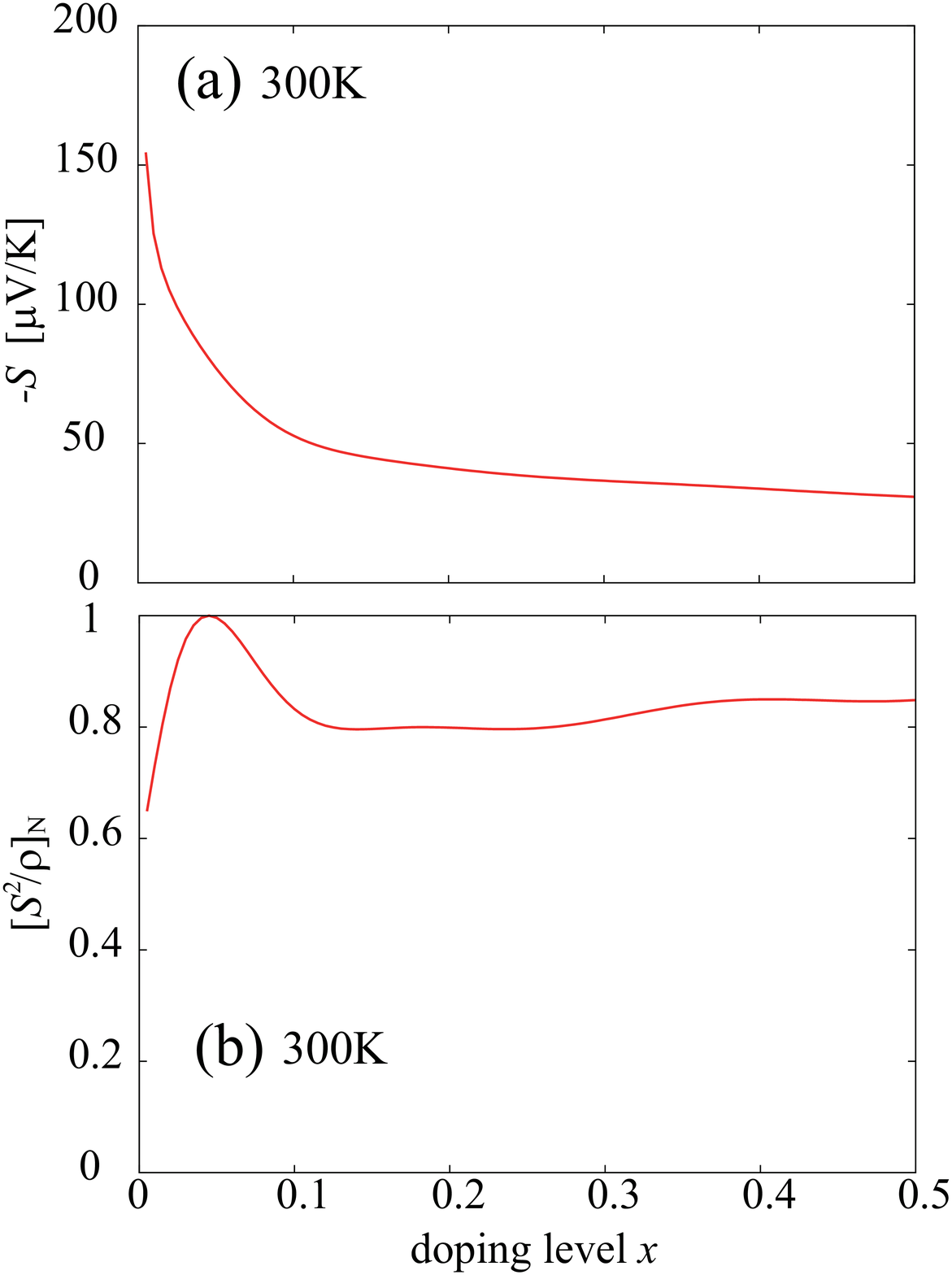}
\caption{(a)The calculated Seebeck coefficient and (b) the normalized 
power factor  at 300K against the doping level $x$.
\label{fig_theory3}
}
\end{figure}

In Fig.\ref{fig_theory3}, 
we show the doping level dependence of the 
Seebeck coefficient and the normalized power factor (normalized by the 
maximum value) at $T=300$K.
This result shows that the Seebeck coefficient keeps relatively large values 
of about $-50 \mu$V/K even at a (hypothetically) 
large doping rate $x \simeq 0.5$, 
so that (assuming a constant quasiparticle life time) 
the power factor is roughly constant. 
This peculiar doping level dependence of the power factor is a 
consequence of the pudding-mold type band.

\subsection{Origin of the pudding-mold type band}
As seen from our theoretical analysis,  
the large Seebeck coefficient in the electron-doped FeAs$_2$ 
is a consequence of the q1D pudding-mold type band.
In this section, we focus on the origin of this peculiar band structure.
For simplicity, 
we first consider a unit cell that contains only one Fe site per unit cell.
This corresponds to expanding (unfolding) the Brillouin zone.
By taking this unit cell, we consider a simplified tight-binding model, 
where only the hoppings 
$t_{z1}$ and $t_{z2}$ along $z$ direction, 
and $t_{xyz}$ to the $(1/2,1/2,1/2)$ position are considered 
(Fig.\ref{fig_theory4}(a), see Fig.\ref{fig_theory4}(b) for the correspondence with the actual lattice structure).
The hoppings in the $x$ and $y$ directions can be neglected for the 
band originating from the $d_{z^2}$ orbital.
In this model, the energy is given as
\begin{eqnarray}
E({\bf k}) = &-&2t_{1z}{\rm cos}(k_z) -2t_{2z}{\rm cos}(2k_z) \nonumber \\ 
 &-&8t_{xyz}{\rm cos}(k_x/2){\rm cos}(k_y/2){\rm cos}(k_z/2).
\end{eqnarray}
When $k_z \simeq \pi$, the third term in the right hand side vanishes 
due to $\cos(\pi/2)=0$, 
and the band dispersion loses $k_x$ and $k_y$ dependence.
This is the explanation of the q1D nature of the 
band structure in the $k_z=\pi$ plane.
We now focus on the band structure along the $k_z$ direction.
When we consider only $t_{1z}$, the band shape is free-electron-like 
at the bottom of the band(Fig.\ref{fig_theory4}(c)).
Adding the distant hopping $t_{2z}=-0.4$, the shape of the band becomes flat near the band edge
around $k_z=\pi$ (corresponding to the Z point), so that
the q1D pudding-mold-type band is formed (Fig.\ref{fig_theory4}(d)).
Further 
considering $t_{xyz}=-0.1$, the band becomes asymmetric  
with respect to $k_z=\pi$ (Fig.\ref{fig_theory4}(e)).

\begin{figure}
\includegraphics[width=8.5cm]{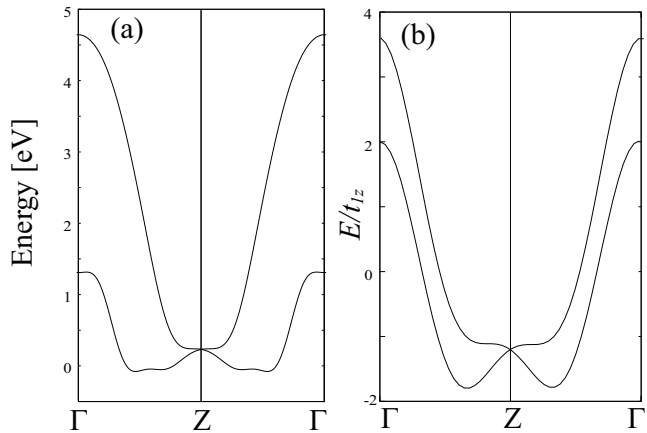}
\caption{(a) The two orbital tight binding model constructed from MLWFs and 
(b) the band structure of Fig.\ref{fig_theory4}(c) shown in the folded 
Brillouin zone.
\label{fig_theory5}
}
\end{figure}

To compare this simple band structure to a more realistic one, 
we construct, again exploiting the MLWF's,  
a two orbital model that considers only the $d_{z^2}$ orbital 
in the original unit cell that contains two Fe.
Although this model does not perfectly reproduce the original 
band structure as in the six orbital model presented in the above,
it still captures the essential features.
In this model, $t_{1z}=-0.65$eV, $t_{2z}/t_{1z}=0.2$, 
$t_{xyz}/t_{1z}=0.3$, and the distant hoppings are also contained.
The band structure of the two orbital model is shown in 
Fig.\ref{fig_theory5}(a) compared with the band structure of 
Fig.\ref{fig_theory4}(c) refolded into the original Brillouin zone 
(Fig.\ref{fig_theory5}(b)).
The band structure of the simplified model is roughly similar to 
that of the two orbital model especially near the band bottom, so that  
the essence of the origin of this peculiar 
band structure can be understood by the simplified model. 
To be precise, the magnitude of $t_{2z}$ and $t_{xyz}$ is different 
between the two models because the simplified model considers 
only three hoppings, while the more realistic two orbital model 
takes into account the distant hoppings as well.
The bottom line here is that the origin of the q1D pudding-mold type 
band is the overlapping feature of the $d_{z^2}$ orbitals, where 
a relatively large second neighbor hopping is present.

\section{Conclusion}
In the present work, we have studied the thermoelectric properties of 
the electron-doped FeAs$_2$. A large Seebeck coefficient is observed 
despite the metallic behavior of the resistivity. First-principles 
band calculation reveals the presence of a q1D pudding-mold type band 
just above the band gap, which mainly originates from the Fe 3dz2 orbital.
orbital. Using a tightbinding model that correctly reproduces this 
band structure, we have calculated the Seebeck coefficient. The 
results are close to the experimental observation, and the 
electron correlation effects are found to be small.
The q1D nature of the pudding-mold type band comes from the 
combination of the $d_{z^2}$ orbital and the lattice structure.
Given the present study, we now have examples of materials 
that possess one-, two-, or three-dimensional 
flat bands that give rise to the large 
Seebeck coefficient in a metallic system.
Therefore, the present findings in the electron-doped 
FeAs$_2$ reinforces the general efficiency of peculiar band shapes 
with partially flat dispersion.

\bibliography{perovskite}

\end{document}